\documentclass[prl,twocolumn,floatfix,superscriptaddress]{revtex4-1}
\usepackage{dcolumn,amsmath}
\usepackage{graphicx}
\usepackage{bm}
\usepackage{hyperref}
\usepackage{xcolor}

\usepackage[utf8]{inputenc}
\usepackage[T1]{fontenc}

\usepackage{tabularx}
\usepackage{braket}
\usepackage{amsfonts}

\begin{document}

\title{Projection population analysis for molecules with heavy and superheavy atoms}

\author{Alexander V. Oleynichenko}
\email{alexvoleynichenko@gmail.com}
\affiliation{Petersburg Nuclear Physics Institute named by B.P.\ Konstantinov of National Research Center ``Kurchatov Institute'' (NRC ``Kurchatov Institute'' - PNPI), 1 Orlova roscha, Gatchina, 188300 Leningrad region, Russia}
\affiliation{Department of Chemistry, M.V. Lomonosov Moscow State University, Leninskie gory 1/3, Moscow, 119991~Russia}
\homepage{http://www.qchem.pnpi.spb.ru}

\author{Andr\'ei V. Zaitsevskii}
\affiliation{Petersburg Nuclear Physics Institute named by B.P.\ Konstantinov of National Research Center ``Kurchatov Institute'' (NRC ``Kurchatov Institute'' - PNPI), 1 Orlova roscha, Gatchina, 188300 Leningrad region, Russia}
\affiliation{Department of Chemistry, M.V. Lomonosov Moscow State University, Leninskie gory 1/3, Moscow, 119991~Russia}

\begin{abstract} 
A new iterative version of population projection analysis is formulated and applied 
to determine relativistic effective atomic configurations of superheavy elements Cn and Fl and their
lighter homologues (Hg and Pb) in the molecules of their fluorides and oxides. The dependence
of the computed populations on the initial reference atomic spinors is completely avoided. 
The difference in population of atomic spinors with the same orbital angular momentum and different total angular momenta
is demonstrated to be essential for understanding the peculiarities of chemical bonding in superheavy
element compounds.
\end{abstract}

\maketitle

%*****************   The Body of the Article:   *************************
\section{Introduction}
In the last two decades transactinide elements with atomic numbers exceeding 110 had been 
synthesized \cite{oganessian09}.
In spite of several successful chemical experiments with these species \cite{eichler13,schadel15}, the bulk of
information on their chemical properties is obtained from electronic structure 
modelling of their compounds.
Due to very strong relativistic effects the bonding pattern in these compounds is rather unusual.
Unfortunately, modern electronic structure theories provide few reliable tools for 
qualitative interpretation of relativistic models in ``chemical'' terms. Up to now,
a straightforward relativistic generalization  of 
Mulliken population analysis \cite{mulliken55} 
remains the most popular tool for estimating the role
of particular atomic spinors in chemical bond formation \cite{pershina02, pershina05, pershina11}.
However, its results are not highly reliable because of their critical dependence on 
the basis set used to discretize the electronic Hamiltonian \cite{jensen} .

%Population analysis can be used as a powerful tool for investigating the chemical bonding in compounds and explaining of experimental facts.
%To be useful, population analysis should give reproducible results and be compatible with wide range of models of electronic structure and
%corresponding quantum chemical methods.

One of the most perspective relativistic population analysis techniques is so-called projection analysis (PA)
proposed by Trond Saue and coworkers \cite{saue96,faegri01,dubillard06} and allowing to calculate effective configurations of atoms
in molecules as sets of fractional occupancies of atomic spinors. The resulting
configurations are much less sensitive to the basis set choice than those obtained via the Mulliken analysis.
It is fully compatible with various tools of relativistic electronic structure calculations, including both all-electron four-component technologies and two-component methods using spin-orbit effective core potentials.
%To the authors' best knowledge, projection analysis in one of only two population analysis techiniques which are able to 
%discern populations of splitted subshells \textbf{(((and Pershina's Mulliken???
%maybe it is more secure to say simply ``Projection analysis is particularly well suited ...''
%)))} such as $p_{1/2}$ and $ p_{3/2}$.
%(another is the <<Atom-in-Compound>> concept, proposed in \cite{titov14,zaitsevskii16}).
Projection analysis is particularly well suited for cases where it is important to discern fractional occupancies of splitted subshells such as $p_{1/2}$ and $ p_{3/2}$.
Therefore the PA technique may be the best choice in the cases where the non-relativistic
and even scalar relativistic description is fully inadequate, including those of  of superheavy element (SHE) compounds.
However, projection analysis employs reference orbitals (spinors) of free atoms, which can depend strongly
on the assumed reference atomic configurations. 
This gives rise to a certain arbitrariness of the computational scheme.

As mentioned above, copernicium (E112) and flerovium (E114) are the most studied in chemical sense among all superheavy elements. The chemistry of elements 112 and 114 is of great interest, since their atoms have quasi-closed-shell electron configurations $7s^2$ for Cn and $7s^2 7p_{1/2}^2$ for Fl with relativistically stabilized valence $7p_{1/2}$ electrons.
This fact should lead to a relative inertness of these elements in most chemical interactions, which has been demonstrated both theoretically \cite{pershina09} and experimentally by thermochromatography on gold surfaces \cite{schadel15}.
Due to their electron configurations Cn and Fl can also possess surprising bonding features in their ``true'' compounds with lighter elements.
Unfortunately, the potential power of the projection analysis for predicting and interpreting the bonding patterns in SHE compounds was not yet widely applied, except hydrides of elements 116 \cite{dubillard06,remigio15} and 117 \cite{saue96} and mixed diatomic molecules TlE117, E113At, E113E117 \cite{faegri01}.
Therefore, the experience of application of projection analysis in this domain is still very scarce and not sufficient to make reliable conclusions about its advantages and drawbacks.

The present paper reports the study of qualitative differences in chemical bonding
in oxides and fluorides of superheavy elements and their lighter homologues,
Cn \emph{vs} Hg and Fl \emph{vs} Pb through evaluating relativistic effective configurations of heavy atoms
in compounds by the projection analysis. The latter technique is slightly modified in order to completely avoid the 
influence of arbitrary choice of atomic reference configurations on the results.
%In the present paper our goals were investigation of bonding features in compounds of Hg, Pb and their superheanvy homologues Cn and Fl
%and improvement of projection analysis technique in order to avoid the arbitrariness arising from the
%dependence on the choice of reference atoms states.

\section{Theory}
The detailed description of the projection analysis technique was given in \cite{dubillard06}.
PA is based on the fact that every molecular spinor $|\psi_i^{MO}\rangle$ can be expanded in the spinors $|\chi^A_k\rangle$ of the constituent atoms, calculated in their own basis sets:
\[ |\psi_i^{MO}\rangle = \sum_A \sum_{k \in A} c_{ki}^A |\chi^A_k\rangle + |\psi^{pol}_i\rangle \]
where $|\psi^{pol}_i\rangle$ stands for the component of $|\psi_i^{MO}\rangle$
outside of the linear span of reference atomic spinors $|\chi^A_k\rangle$.

%Since $|\psi^A_p\rangle$ are not guaranteed to fully span the molecular \textbf{spinors} the expansion includes
%their orthogonal complement $|\psi^{pol}_i\rangle$, denoted the polarization contribution.

The expansion coefficients $c_{ki}^A$ are found by solving the set of linear equations, obtained by projection:
\[ \langle \chi_k^B | \psi_i^{MO} \rangle = \sum_A \sum_{q \in A} \langle \chi_k^B | \chi_q^A \rangle c_{qi}^A \]

In fact, we reexpand molecular spinors in the basis of atomic ones $|\chi^A_k\rangle$.
Then fractional occupancies $N_k^A$ of the new basis orbitals $|\chi^A_k\rangle$ are calculated similarly to Mulliken analysis:
\[N_k^A = (\mathbb{P}^{at}\mathbb{S}^{at})_{kk}^A = \sum_q^{all AOs} \sum_i^{occ MOs} c_{ki}^A c_{iq}^{*} \langle \chi_q | \chi_k^A \rangle \]

Note that in general the sum of fractional occupancies should not coincide with the total number of electrons $N$.
Let us introduce the number $N_{pol}$ of electrons which are ``lost'' at the first projection step:
\[ N_{pol} = N_{elec} - \sum_{A} \sum_{k \in A} N_k^A \]
Obviously, if $N_{pol}$ is large the analysis becomes senseless because a significant
fraction of electron density is not assigned to any atom.

In order to completely eliminate polarization contribution molecular spinors (or pseudospinors) can be transformed to "intrinsic" atomic spinorsc \cite{knizia13}. However, it slightly distorts clear physical meaning of effective atomic configuration, since the "intrinsic" atomic spinors do not possess any definite angular momentum with respect to the nucleus of the corresponding atom.

As has been mentioned above, the resulting fractional occupancies $N^A_k$ depend on the particular choice of atomic spinors $|\chi^A_k\rangle$,
which are in turn defined by the configurations of reference atoms chosen with an inavoidable degree of arbitrariness. 
Furthermore, certain spinors which are not occupied in ground-state atoms but play an important role in the bond formation are sometimes hardly usable for PA. For instance, reference atomic spinors obtained in conventional Hartree-Fock method (HF) or Kohn-Sham DFT method with integral spinor occupancies are as diffuse as allows the chosen basis set. 
Moreover, in some cases they even can have no physical meaning. This observation is particularly important since several heavy atoms are not capable to attach electron and thus have no electron affinity (for example, Hg and both Cn and Fl \cite{borschevsky09}).
Conventional PA underestimates populations of such spurious atomic spinors, whereas their compact counterparts are strongly involved in bond formation and should be taken into account while considering bonding features.

%As a result, PA populations of such virtual atomic spinors are underestimated.
%For example, atomic orbitals of Hg, calculated for the $6s^2 6p_{1/2}^0 6p_{3/2}^0$ configuration, will give fundamentally wrong
%fractional occupancies of $6p_{1/2}$ and $6p_{3/2}$ orbitals of Hg in its compounds.

To avoid this shortcoming at least partially, we proposed an iterative version of the PA approach, based 
on the use of fractional-occupancy reference atomic configurations.
The idea is very simple: the occupancies, obtained via the projection analysis at the first step,
are used as input fractional occupancies for the new atomic calculation. The obtained atomic spinors are used as a new reference atomic in PA again. This loop is repeated until convergence. This procedure is shown to converge very fast and reduces significantly polarization contributions (see below).

%As a result of described procedure, low-lying virtual atomic spinors become more spatially compact.

The proposed method is suitable at least for atoms bearing positive net charges. Otherwise the atomic calculations for negative-charged ion must be carried out, this again can lead to unphysical and spatially blurred atomic spinors which cannot be corrected by introduction of fractional occupancies. In this case proposed iterative procedure does not converge and exhibits strong oscillations of subshell effective occupancies.

\section{Computational details}
The iterative projection analysis (IPA) was applied to the molecules of oxides and difluorides
of superheavy elements Cn, Fl and their lighter homologues (Hg and Pb respectively).
All calculations were carried out by the two-component Kohn-Sham method with PBE0 functional \cite{adamo99},
implemented in the DIRAC code \cite{dirac15}.
We replaced core electrons with spin-orbit
semilocal relativistic effective core potentials by Mosyagin et al \cite{mosyagin10} (60-e RECPs for Hg, Pb and 92-e RECPs for Cn, Fl).
Valence and subvalence spinors of heavy atoms were represented using flexible uncontracted basis set from \cite{pnpi_site}.
We also used cc-pVTZ \cite{dunning89} basis set for oxygen and fluorine.
Before running population analysis the geometries of $\rm MF_2$ and $\rm MO$ (M = Hg, Cn, Pb, Fl) were optimized (see Table 1). For performing one-step projection analysis we also used the code included in the DIRAC package.

In order to find trends in variations of typical bond strength in pairs Hg -- Cn and Pb -- Fl we also
evaluated dissociation energies of all studies molecules. Since Kramers-restricted configuration-averaged approach is not well suited
for DFT calculations of open-shell species, we used unresticted two-component RDFT code by Christoph van W\"{u}llen \cite{wullen10}
to calculate open-shell atomic ground states.

\section{Results and discussion}

\begin{table*}[t!]
\caption{\label{tab:table1} \scriptsize
Effective relativistic configurations of heavy atom M according to the iterative projection analysis, equilibrium geometries and bond energies $E_{{\rm M-X}}$ in difluoride and oxide molecules.
$E_{{\rm M-X}}$, X=F, O, are the adiabatic energies for  1/2 MF$_2 \rightarrow 1/2$ M + F
and $\rm MO \rightarrow $M + O reactions.
}
\begin{ruledtabular}
\begin{tabular}{ccccccccc}
         & \multicolumn{5}{c}{Subshell occupancies} & Bond  & Valence   & $ E_{{\rm M-X}}$, \\
Molecule & $ s_{1/2}$ & $ p_{1/2}$ & $ p_{3/2}$ & $ d_{3/2}$ & $ d_{5/2}$ & length, A & angle, deg &  eV \\
\hline
\\
HgO		&	0.65	&	0.05	&	0.07	&	3.96	&	5.90	&	1.875   &	       & 0.06  \\
CnO		&	1.71	&	0.33	&	0.10	&	3.94	&	5.45	&	1.861   &	      & 0.23  \\
HgF2	&	0.94	&	0.11	&	0.15	&	3.92	&	5.80	&	1.902   &	180.0   & 2.75  \\
CnF2	&	1.50	&	0.16	&	0.10	&	3.88	&	5.46	&	1.931   &	180.0   & 2.17  \\
\\
PbO		&	1.88	&	0.69	&	0.63	&	3.99	&	5.99	&	1.898	&	       & 3.69  \\
FlO		&	1.93	&	1.18	&	0.28	&	3.98	&	5.97	&	2.038	&	       & 0.83  \\
PbF2	&	1.88	&	0.39	&	0.46	&	4.00	&	5.99	&	2.023	&	95.6    & 4.15  \\
FlF2	&	1.95	&	0.63	&	0.28	&	3.99	&	5.98	&	2.154	&	96.9    & 2.17  \\
\\
\end{tabular}
\end{ruledtabular}
\end{table*}

Equilibrium geometry parameters (bond lengths and angles) and effective configurations of heavy atoms in the studied molecules are summarized in Table~\ref{tab:table1}.

We found that the results of the iterative projection analysis are fully independent on the initial reference atomic orbitals
and iterations converge very rapidly even with nearly senseless starting atomic spinors (e.g. those of $\rm Cn^{20+}$ ion; we recall that for Cn we use 92-e RECP).
%Thus, iterative PA technique is fully insensitive to the choice of initial subshell fractional occupancies of reference atoms.

Furthermore, during the iterations the polarization contribution decreases for every tried initial guess.
We also studied the effect of the transformation of MOs to IAOs on results of projection analysis and found that
it slightly affects the results (order of 0.01 of a atomic charge unit). Hence transformation to IAOs is not critical
and can easily be avoided by an iterative procedure.

\begin{figure}[t]
\begin{center}
  \includegraphics[width=\columnwidth]{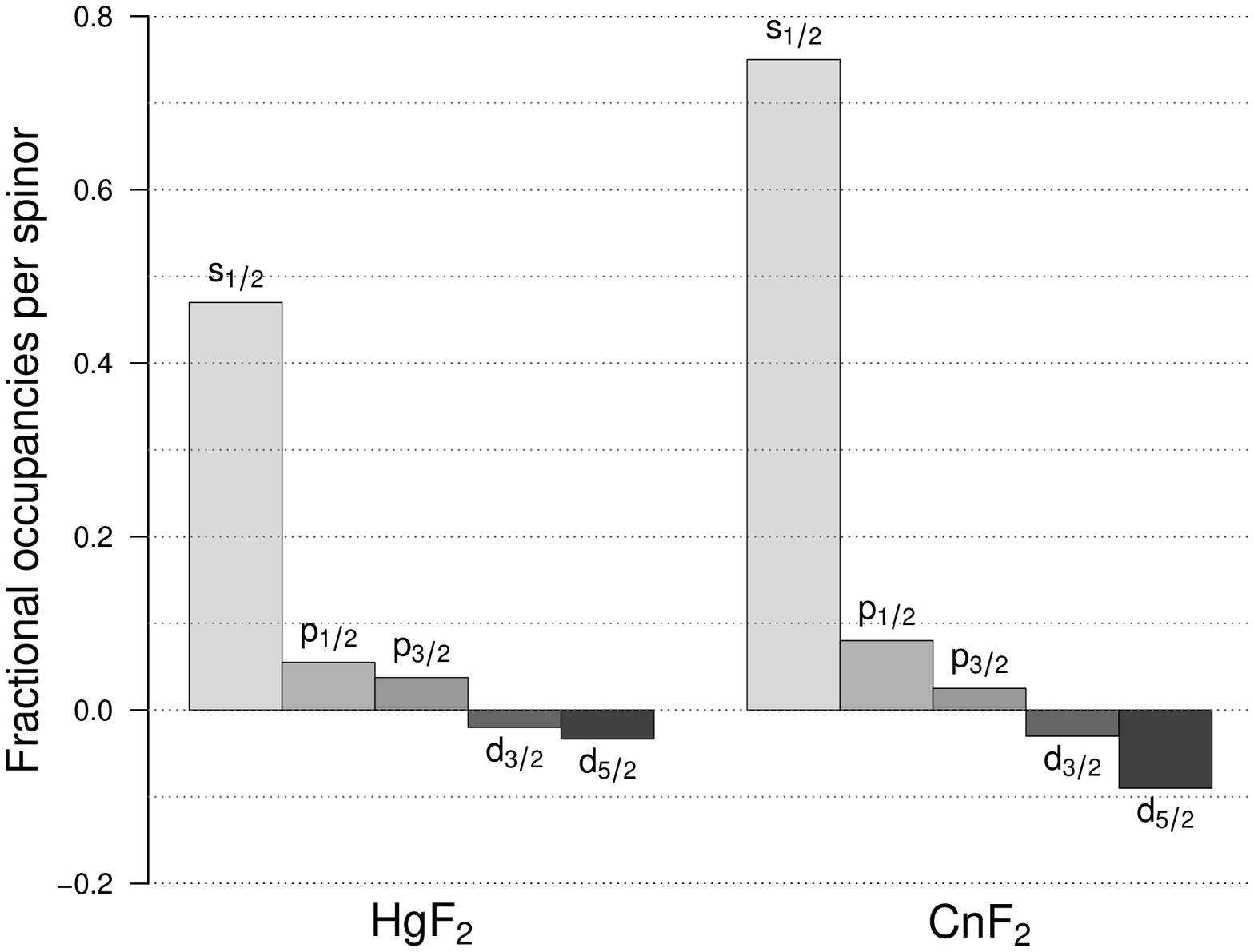}
  \includegraphics[width=\columnwidth]{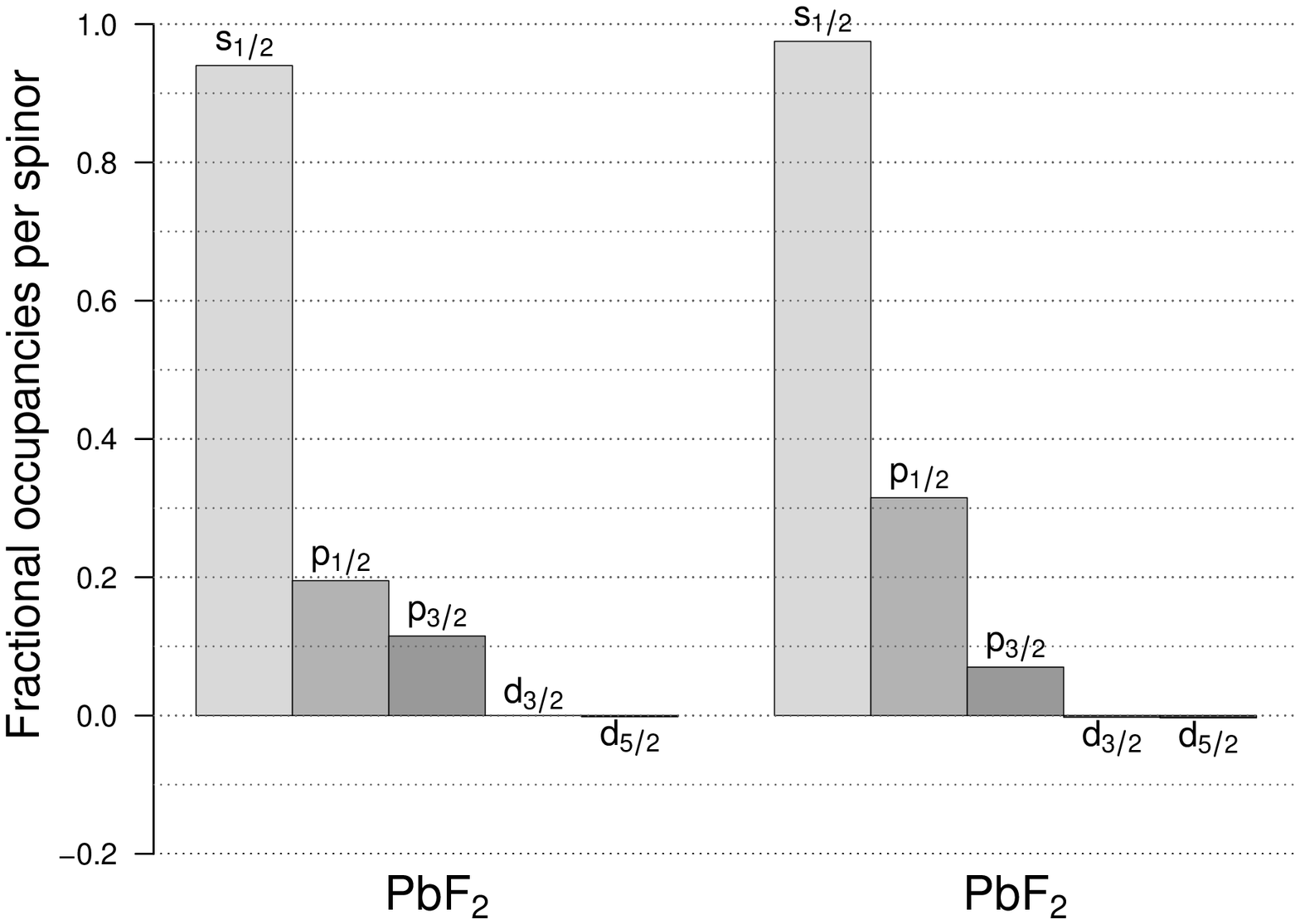}
\end{center}
  \caption{\scriptsize Effective subshell ocupancies \emph{per spinor} for heavy atoms in difluorides $\rm MF_2$ (M = Hg, Cn, Pb, Fl),
  obtained via the iterative projection analysis. Values are given relative to the configuration
  $(n-1)s^2(n-1)p_{1/2}^2(n-1)p_{3/2}^4d_{3/2}^2(n-1)d_{5/2}^4$ (filled subvalence shells),
  negative $d$-occupancies correspond to subvalence $d$-shell holes. }
\end{figure}

%In order to clarify the features of chemical bonding in molecular compounds of superheavy elements 
%we have applied the new iterative PA technique to oxides and diflourides of Hg, Cn, Pb and Fl.

To interpret our results in intuitive ``chemical'' terms,  let us recall that in the non-relativistic
 (and scalar-relativistic) models the formation of one or two covalent bond by an atom in a closed shell
 molecule implies the presence of one or two unpaired electron above the closed shell in the free atom  In the idealized case
of purely covalent bonding, the overall populations of involved atomic orbitals remains the same
 upon the bond formation. If the atomic subshell population approaches that of the filled shell,
 the same holds for hole populations. Therefore the deviation of the atomic $l$-subshell occupancy 
 from  0 (for low populations) or from $l(2l+1)$ (for high populations) in a closed-shell molecule
 is related the number of single covalent bonds formed by the atom. 
 It is natural to suppose that this deviation should correlate with the covalent
 contribution to chemical bonding also in the case of fractional populations.

These considerations, however, should be modified to the relativistic case. Let us recall that the 
atomic two-component spinors are admixtures of spin-up and spin-down components. For example,  $p$ spinors ($l=1$ with the total spin projection $m_j > 0$  have the form
\[ p_{1/2,1/2} = \left( \begin{array}{c} \frac{1}{\sqrt{3}} p_0 \\ \sqrt{\frac{2}{3}} p_1 \end{array} \right) \]
\[ p_{3/2,1/2} = \left( \begin{array}{c} \sqrt{\frac{2}{3}} p_0^{\prime} \\ \frac{1}{\sqrt{3}} p_1^{\prime} \end{array} \right) \quad\quad
   p_{3/2,3/2} = \left( \begin{array}{c} p_1^{\prime} \\ 0 \end{array} \right)
\]
where the radial parts of $p^{\prime}$ and $p$ are different. Provided that we have a single
electron on the $p$-subshell, the formation of a true $\sigma$-bond along an
interatomic axis (let us call it $z$) implies the participation of an electron on
the $p\sigma{}$ (or, the same, $l=0$) component. However, neither any of the  $p_{1/2}$ spinors  nor
any combination of these spinors is dominated by this component. The same holds
for the $p_{3/2}$ spinors; to get a pure $p\sigma{}$ function, $p$ spinors with different $j$
are to be combined~\cite{lee04}. Such mixing is not efficient when the difference of energies and spatial distributions of the $j$-subshells with the same $l$ is large, as occurs in superheavy elements. 

%Note that only the spinors with the same occupancies can be mixed.
%, population of the ``mixed'' (or ``hybrid'') spinor is  twice the 
%occupancy of less occupied spinor in the pair $| l, l-1/2 \rangle$, $| l, l+1/2 \rangle$. Thus effective amount of ``unpaired'' electrons taking part in $\sigma$-bond formation is equal to occupancy of less occupied spinor plus the difference in occupations of spinors with the same $l$ 
%divided by the coefficient of the $z$-directed component in the spherical spinor expression.
%
%This outline leads to conclusion that 
A large difference between the populations (per spinor) of the shells  with the same $l$ and
different $j$
indicates inefficient mixing and therefore weakening of $\sigma$-bonds. Note that the above considerations can be extended %straightforward
 to other types of chemical bonds (but $\sigma$ bonds are the strongest). 
 %In fact bond energy will be determined by percent of $\sigma$-bonding, and therefore by occupations of $s$-, $p_{1/2}$- and $p_{3/2}$ subshells (not $d$).

These arguments can be used for interpretation of results obtained via the iterative PA technique. In case of compounds of group 12 elements (Hg and Cn) chemical bonds are formed mainly by $s$ electrons (See Figure 1). We may assert that the covalent component of chemical bonds is stronger in $\rm HgF_2$ than in $\rm CnF_2$ since the fractional occupancy of the $s$-shell approaches 1. 
%Since the contributions of $p$ electrons are small and comparable and then $p$-spinors are not very important in bonding, we may predict Hg--F bond to be stronger than Cn--F only on the basis of $s$-occupancies. 
Furthermore, though hole population of the Cn $d_{5/2}$ subshell is significant, it does not
strongly  contribute to $\sigma$-bonding since the very small hole population of $d_{3/2}$ indicates 
the inefficiency of $d_{5/2}-d_{3/2}$ mixing required to form $d\sigma$ components. However, $d$-subshells seems not to be as inert as in the case of Hg where they are nearly filled, that characterizes Cn as a real transition element.

This qualitative picture agrees with the bond energy estimates (~2.8 eV for Hg--F bond and ~2.2 eV for Cn--F one). The smallness of Hg-O and Cn-O bond energies block the possibility
of searching the correlations between the bonding pattern and
bond strength in this case.
%Relative inertness of Cn can be easily explained by relativistic contraction of $s$- and $p_{1/2}$-%subshells in passing from from Hg to Cn. 
\begin{figure}[t]
\begin{center}
  \includegraphics[width=\columnwidth]{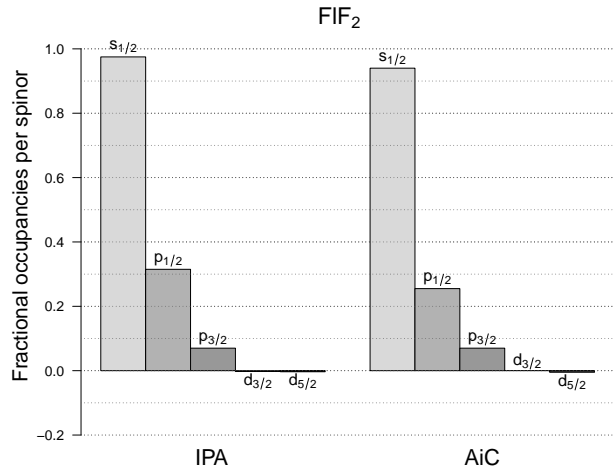}
\end{center}
  \caption{\scriptsize Effective subshell occupancies of Fl atom in $\rm FlF_2$, obtained via iterative PA (left) and AiC \cite{romanov16} (right). }
\end{figure}

A different bonding pattern is observed for the compounds of Pb and Fl. In this case the results of the iterative projection analysis 
are consistent with the simple chemical intuition: subvalence $d$- and valence $s$-shells are actually quasi inactive in bond formation, and these two elements exhibit typical $p$-element behavior. Again, from Pb to Fl, mixing between the subshells with the same $l$ and  becomes less efficient, and the significantly populated of Fl $p_{1/2}$ subshell in FlO and $\rm FlF_2$ cannot greatly contribute to covalent bonding due to the imbalance of $p_{1/2}$ and $p_{3/2}$ occupancies. 
This leads to a weakening of bonds involving flerovium atom with respect to those formed by Pb.
This conclusion is also consistent with decreasing of calculated M--F bond energy from 4.2 eV for $\rm PbF_2$ to 2.2 eV for $\rm FlF_2$.
%In addition, in the basis of small fractional occupancies of $p$ subshells we may conclude that chemical bonds have a significant ionic character (maybe even more ionic than in $\rm HgF_2$ and $\rm CnF_2$).
Essentially the same pattern is observed for the corresponding oxides. 
%However, bonds with oxygen seem to have more covalent character than with fluorine.

Finally, we should point out that all results obtained with iterative PA qualitatively coincide with the predictions of the AiC analysis \cite{titov14,zaitsevskii16,romanov16} (see Figure 2). Both PA and AiC approaches give grounds to assert that bonds in $\rm PbF_2$ and $\rm FlF_2$ are formed entirely by $p$-electrons.

\section{Conclusions}
To summarise, the new iterative approach to projection population analysis was proposed
and shown to be rapidly converging and fully insensitive to the initial choice of reference atomic configuration. Based on conventional projection analysis technique,
it can be easily used on the top of any quantum chemistry code implementing fractional-occupancy two- or four-component relativistic Hartree-Fock or DFT methods.

The application of iterative projection analysis to the molecules of Hg, Cn, Pb and Fl
oxides and difluorides gives some insight into the peculiarities of chemical bonding
in superheavy element compounds. The decrease of the covalent component of bonds in Cn compound with
respect to their Hg-containing counterparts is related to the increase of the $s$-subshell
population approaching that of the closed shell. Furthermore, the large difference of hole
populations of $d_{3/2}$ and $d_{5/2}$ subshells prevents
their admixture necessary for strong $\sigma$-bonding.
A similar imbalance of $p_{1/2}$ and $p_{3/2}$ occupancies results in weakening of covalent bonding
in Fl compounds. The separate determination of effective populations for the subshells with the same
orbital angular momentum but different total angular momenta is thus crucial for interpreting the bonding pattern on superheavy element compounds.

This study was supported by the Russian Science Foundation (grant no. 14-31-00022).


\begin{thebibliography}{99}
\bibitem{oganessian09} Y. Oganessian. Heavy element research at FLNR (Dubna). Eur. Phys. J. A. {\bf 42}, 361-367 (2009).
\bibitem{eichler13}    R. Eichler. First foot prints of chemistry on the shore of the Island of Superheavy Elements. J. Phys. Conf. Ser. {\bf 420}, 012003 (2013).
\bibitem{schadel15}    M. Sch\"{a}del. Chemistry of the superheavy elements. Phil. Trans. R. Soc. {\bf 373}, 20140191 (2015).
\bibitem{mulliken55}   R. S. Mulliken. Electronic Population Analysis on LCAO-MO Molecular Wave Functions I. J. Chem. Phys. {\bf 23}, 1833 (1955).
\bibitem{pershina02}   V. Pershina, T. Bastug, T. Jacob, B. Fricke, S. Varga. Intermetallic compounds of the heaviest elements: the electronic structure and bonding of dimers of element 112 and its homolog Hg. Chem. Phys. Lett. {\bf 365}, 176-183 (2002).
\bibitem{pershina05}   V. Pershina, T. Bastug. Relativistic effects on experimentally studied gas-phase properties of the heaviest elements. J. Chem. Phys. {\bf 311}, 139-150 (2005).
\bibitem{pershina11}   V. Pershina, A. Borschevsky, J. Anton. Fully relativistic study of intermetallic dimers of group-1 elements K through element 119 and prediction of their adsorption on noble metal surfaces. J. Chem. Phys. {\bf 395}, 87-94 (2012).
\bibitem{jensen}       F. Jensen. {\it Introduction to Computational Chemistry}. John Wiley \& Sons (2007).
\bibitem{saue96}       T. Saue, K. Faegri, O. Gropen. Relativistic effects on the bonding of heavy and superheavy hydrogen halides. Chem. Phys. Lett. {\bf 263}, 360-366 (1996).
\bibitem{faegri01}     K. Faegri, T. Saue. Diatomic molecules between very heavy elements of group 13 and group 17: A study of relativistic effects on bonding. J. Chem. Phys. {\bf 115}, 2456 (2001).
\bibitem{dubillard06}  S. Dubillard, J.-B. Rota, T. Saue. Bonding analysis using localized relativistic orbitals: Water, the ultrarelativistic case and the heavy homologues $\rm H_2 X$ (X = Te, Po, eka-Po). J. Chem. Phys. {\bf 124}, 154307 (2006).
\bibitem{pershina09}   V. Pershina, J. Anton, T. Jacob. Theoretical predictions of adsorption behavior of elements 112 and 114 and their homologs Hg and Pb. J. Chem. Phys. {\bf 131}, 084713--084718 (2009).

%\bibitem{kullie16}     O. Kullie. Density Functional study of Covalency in the Trihalides of Lutetium and Lawrencium. arXiv:1608.04703 [physics.chem-ph].
%\bibitem{south16}      C. South, A. Shee, D. Mukherjee, A. K. Wilson, T. Saue. 4-Component relativistic calculations of $\rm L_3$ ionization and excitations for the isoelectronic species $\rm UO_2^{2+}$, $\rm OUN^{+}$ and $\rm UN_2$. Phys. Chem. Chem. Phys. {\bf 18}, 21010-21023 (2016).
\bibitem{remigio15}    R. Di Remigio, R. Bast, L. Frediani, T. Saue. 4-Component Relativistic Calculations in Solution with the Polarizable Continuum Model of Solvation: Theory, Implementation and Application to the Group 16 Dihydrides $\rm H_2X$ (X = O, S, Se, Te, Po). J. Phys. Chem. A. {\bf 119}, 5061-5077 (2015).
%\bibitem{knecht11}     S. Knecht, S. Fux, R. van Meer, L. Visscher, M. Reiher, T. Saue. Mossbauer spectroscopy for heavy elements: a relativistic benchmark study of mercury. Theor. Chem. Acc. {\bf 129}, 631-650 (2011).
\bibitem{knizia13}     G. Knizia. Intrinsic atomic orbitals: An unbiased bridge between quantum theory and chemical concepts. J. Chem. Theory Comput. {\bf 9}, 4834-4843 (2013).
\bibitem{borschevsky09} A. Borschevsky, V. Pershina, E. Eliav, U. Kaldor. Electron affinity of element 114, with comparison to Sn and Pb. Chem. Phys. Lett. {\bf 480}, 49 (2009).
\bibitem{adamo99}      C. Adamo, V. Barone. Toward reliable density functional methods without adjustable parameters: The PBE0 model. J. Chem. Phys. {\bf 110}, 6158 (1999).
\bibitem{dirac15}
DIRAC, a relativistic ab initio electronic structure program, Release DIRAC15 (2015), written by R. Bast, T. Saue, L. Visscher, and H. J. Aa. Jensen, with contributions from V. Bakken, K. G. Dyall, S. Dubillard, U. Ekstroem, E. Eliav, T. Enevoldsen, E. Fasshauer, T. Fleig, O. Fossgaard, A. S. P. Gomes, T. Helgaker, J. Henriksson, M. Ilias, Ch. R. Jacob, S. Knecht, S. Komorovsky, O. Kullie, J. K. Laerdahl, C. V. Larsen, Y. S. Lee, H. S. Nataraj, M. K. Nayak, P. Norman, G. Olejniczak, J. Olsen, Y. C. Park, J. K. Pedersen, M. Pernpointner, R. Di Remigio, K. Ruud, P. Salek, B. Schimmelpfennig, J. Sikkema, A. J. Thorvaldsen, J. Thyssen, J. van Stralen, S. Villaume, O. Visser, T. Winther, and S. Yamamoto (see http://www.diracprogram.org). 
\bibitem{mosyagin10}   N. S. Mosyagin, A. Zaitsevskii, A. V. Titov. Shape-consistent relativistic effective potentials of small atomic core. Int. Rev. At. Mol. Phys. {\bf 1}, 63-72 (2010).
\bibitem{pnpi_site}    http://www.qchem.pnpi.spb.ru/recp.html.
\bibitem{dunning89}    T. H. Dunning. Gaussian basis sets for use in correlated molecular calculations. I. The atoms boron through neon and hydrogen. J. Chem. Phys. {\bf 90}, 1007 (1989).
\bibitem{wullen10}     C. van W\"{u}llen. A Quasirelativistic Two-component Density Functional and Hartree-Fock Program. J. Phys. Chem. {\bf 224}, 413 (2010).
%\bibitem{eliav95}      E. Eliav, U. Kaldor. Transition energies of mercury and ekamercury (element 112) by the relativistic coupled-cluster method.Phys. Rev. A. {\bf 52}, 2765 (1995).
\bibitem{lee04}        Y. S. Lee. Two-component relativistic effective core potential calculations for molecules. Theor. Comput. Chem. {\bf 14}, 352--416 (2004).
\bibitem{titov14}      A. V. Titov, Yu. V. Lomachuk, L. V. Skripnikov. Concept of effective states of atoms in compounds to describe properties determined by the densities of valence electrons in atomic cores. Phys. Rev. A. {\bf 90}, 052522 (2014).
\bibitem{zaitsevskii16} A. Zaitsevskii, L. V. Skripnikov, A. V. Titov. Chemical bonding and effective atomic states of actinides in higher oxide molecules. Mendeleev Commun. {\bf 26}, 307-308 (2016).
\bibitem{romanov16}    A. Zaitsevskii, S. A. Romanov, A. Oleynichenko, L. V. Skripnikov, A. V. Titov. To be published.

%\bibitem{pipek89}
%Pipek\,J., Mezey\,P.\,G. {\it A fast intrinsic localization procedure applicable for ab initio and
%semiempirical linear combination of atomic orbital wave functions}, Journal of Chemical Physics {\bf 90}, 4916 (1989).
%\bibitem{reed85}
%Reed\,A.\,E., Weinstock\,R.\,B. Weinhold F. {\it Natural population analysis}, Journal of Chemical Physics {\bf 83}, 735 (1985).
\end{thebibliography}
\end{document}